\documentclass[twocolumn,preprintnumbers,amsmath,amssymb,prb]{revtex4}
\usepackage{graphicx}% Include figure files
\usepackage{dcolumn}% Align table columns on decimal point
\usepackage{bm}% bold math
\usepackage{color}
\begin{document}

\title{Magnetism driven by fluctuations and frustration in synthetic triangular antiferromagnets with ultracold fermions in optical lattices}

\author{Daisuke Yamamoto,$^1$ Giacomo Marmorini,$^{2}$ Masahiro Tabata,$^1$ Kazuki Sakakura,$^1$ and Ippei Danshita$^{3}$}
\affiliation{$^1$Department of Physics and Mathematics, Aoyama Gakuin University, Sagamihara, Kanagawa 252-5258, Japan}
\affiliation{$^2$Research and Education Center for Natural Sciences, Keio University, Kanagawa 223-8521, Japan}
\affiliation{$^3$Department of Physics, Kindai University, Higashi-Osaka, Osaka 577-8502, Japan}
\date{\today}% It is always \today, today,
             % but any date may be explicitly specified
\begin{abstract}
Quantum simulators based on cold atomic gases can provide an ideal platform to study the microscopic mechanisms behind intriguing properties of solid materials and further explore novel exotic phenomena inaccessible by chemical synthesis. Here we propose and theoretically analyze a coherently coupled binary mixture of Fermi atoms in a triangular optical lattice as a promising realization of synthetic frustrated antiferromagnets. We perform a cluster mean-field plus scaling analysis to show that the ground state exhibits several nontrivial magnetic phases and a novel spin reorientation transition caused by the quantum order-by-disorder mechanism. Moreover, we find from {Monte Carlo} simulations that thermal fluctuations induce {an unexpected coexistence of Berezinskii-Kosterlitz-Thouless physics and long-range order in different correlators. } 
These predictions, besides being {relevant to present and future experiments on} triangular antiferromagnetic materials, can be tested in the laboratory with the combination of the currently available techniques for cold atoms.
\end{abstract}
\maketitle
%\emph{Introduction.}--- 

Quantum simulations of solid materials with more flexible systems based on cold atomic gases have attracted {increasing} attention as the third platform to complement the conventional theoretical and experimental studies~\cite{lewenstein-07,bloch-12,gross-17}. In particular, a binary mixture of Fermi atoms in an optical lattice~\cite{greif-13,hart-15,greif-15,parsons-16,boll-16,cheuk-16,mazurenko-17,hilker-17,brown-17,ozawa-18} is a promising candidate to simulate the Hubbard model that describes strongly correlated lattice electrons and is believed to capture a wide variety of intriguing phenomena including the Mott insulator transition, quantum magnetism~\cite{auerbach-94}, and high-temperature superconductivity~\cite{lee-06}.

Although Fermi gases in optical lattices used to suffer from a technical difficulty in cooling to the regime {of magnetic ordering} where many-body spin correlation appears on stage, that is being overcome by recent innovative experimental~\cite{taie-12,hart-15,mazurenko-17} and theoretical~\cite{bernier-09,mathy-12,kantian-18,goto-17} efforts. There has also been rapid development on detection techniques of quantum states, especially imaging individual atoms by the quantum gas microscope~\cite{parsons-16,boll-16,cheuk-16,mazurenko-17,hilker-17,brown-17}. Owing to them, the observations of dimerized~\cite{greif-13}, short-ranged~\cite{hart-15,greif-15,parsons-16,boll-16,cheuk-16,brown-17}, long-ranged~\cite{mazurenko-17}, string~\cite{hilker-17} and SU($\mathcal{N}$)~\cite{ozawa-18} spin correlations have been achieved in the past few years. These important progresses have opened  a new chapter in the cold-atom quantum simulation to study the broad field of strongly correlated magnetism.

Here we make a proposal and provide the necessary theoretical modeling in order to push the boundaries of cold-atom simulation towards the study of a more challenging issue, namely, quantum frustrated magnetism. Although many intriguing phenomena, such as quantum spin liquids~\cite{balents-10}, multipolar or multi-$Q$ orders~\cite{shannon-06,zhitomirsky-10,li-16,kamiya-14,giacomo-14}, {and unconventional quantum criticalities~\cite{vojta-18}}, have been suggested for frustrated systems, {complete understanding} of the mechanism and the properties often remains out of reach because of the difficulty in preparing ideal model materials {with} chemical synthesis and of the lack of suitable numerical methods {for analyzing} the model Hamiltonian. Motivated by the success of the Greiner group {at Harvard University} in simulating a long-range (unfrustrated) N\'eel order with a square-lattice cold-atom simulator~\cite{mazurenko-17}, we suggest a synthetic Hubbard system of binary Fermi gases in a triangular optical lattice for studying the frustrated magnetism deeply connected to the ongoing experiments on quasi-two-dimensional (quasi-2D) layered materials of quantum triangular-lattice antiferromagnets (TLAFs)~\cite{ishii-11,shirata-12,zhou-12,susuki-13,lee-14-2,yamamoto-15,yokota-14,koutroulakis-15,ito-17,paddison-17,cui-18}. {We take into consideration the physically interesting parameter space defined by the spin-exchange anisotropy (parameterized by $J/J_z$) of easy-axis type and the magnetic field applied perpendicular to the anisotropy axis,} which can be created in cold-atom experiments by state-dependent optical lattices~\cite{mandel-03,lee-07,jotzu-15} and by coherent couplings between two hyperfine states with Raman laser beams or radio-frequency field created in an atom chip~\cite{goldman-10,anderson-13}.

Although TLAFs are prototypical model systems for geometrical frustration, there remain unanswered fundamental questions regarding nontrivial excitation spectra~\cite{ito-17,paddison-17}, various magnetization processes~\cite{fortune-09,susuki-13,lee-14,yamamoto-15}, the emergence and {properties}  of quantum spin liquids~\cite{paddison-17,shimizu-03,li-15}, etc. {From the standpoint of real materials}, only very few available compounds have an equilateral triangular-lattice structure with quantum spin $S=1/2$~\cite{susuki-13,lee-14,yamamoto-15,shirata-12,zhou-12,koutroulakis-15,yokota-14,cui-18} and they typically do not fall into the range of easy-axis anisotropy $0<J/J_z\leq 1$ (with the possible exception of Ba$_3$CoNb$_2$O$_9$, according to {Ref.~\onlinecite{yokota-14}}; it is also worth mentioning Rb$_4$Mn(MoO$_4$)$_3$~\cite{ishii-11} and Ba$_3$MnNb$_2$O$_9$~\cite{lee-14-2}, which are however $S=5/2$). 
Therefore, the quantum simulations of TLAFs will be a significant step towards one of the fundamental goals of cold-atom quantum simulators, namely to obtain a deeper understanding of solid-state physics together with the possibility of testing it over a {wider} parameter range.

On the theoretical side, the interplay of quantum or thermal fluctuations and geometrical frustration in {our proposed system with no $U$(1) spin-rotational symmetry} is an open problem. {The lack of the symmetry} inhibits {the decomposition of the Hilbert space into fixed spin subspaces. This makes numerical calculations with, e.g.,  exact diagonalization and density matrix renormalization group (DMRG) even more difficult, since they have a serious limitation on the tractable system size in two dimensions (2D) or higher.} Besides, the quantum {Monte Carlo} (QMC) method suffers from the notorious minus-sign problem~\cite{suzuki-93} for frustrated systems.

To overcome this numerical challenge, here we develop {a new framework of the cluster mean-field method} combined with a scaling analysis (CMF+S)~\cite{yamamoto-12-2,yamamoto-14,yamamoto-17} by employing the 2D DMRG algorithm as a solver of cluster problems with mean-field boundary conditions. This enables us to discuss the ground state of the quantum frustrated system in a quantitative way. We find that a particular quantum selection of the ground state gives rise to a novel gradual reorientation transition of three-sublattice magnetic orders, in addition to the stabilization of nonclassical phases in a wide region of the quantum phase diagram. 
{Furthermore, performing classical Monte Carlo simulations to investigate thermal fluctuations, we show that the paramagnetic transition at finite temperatures exhibits a two-step behavior through an intermediate phase with an unexpected coexistence of Berezinskii-Kosterlitz-Thouless (BKT) physics and long-range order (LRO) in different correlators. } 
Our theoretical phase diagrams are meant to play a role in the  two-way relationship which characterizes this early stage of the quantum simulation of frustrated physics: on the one hand, they act as a benchmark for future cold-atom experiments; on the other hand, they can and must be checked with the help of the quantum simulator. 
\\ \\    
{\bf \large Results}\\
{\bf Synthetic spin-1/2 XXZ model with a transverse magnetic field.} 
The system of coherently coupled Fermi mixtures {of} two hyperfine states ($\sigma=\uparrow, \downarrow$) is described by the Hamiltonian
\begin{eqnarray}
\hat{\mathcal{H}}&=&-\sum_{\langle i,j\rangle,\sigma}t_{\sigma}\left(\hat{c}_{i\sigma}^\dagger\hat{c}_{j\sigma}+{\rm H.c.}\right)+U\sum_{i}\hat{n}_{i\uparrow}\hat{n}_{i\downarrow}\nonumber\\&&+\frac{\Omega}{2}\sum_i\left( \hat{c}_{i\uparrow}^\dagger\hat{c}_{i\downarrow}+\hat{c}_{i\downarrow}^\dagger\hat{c}_{i\uparrow}\right). \label{hamiltonian}
\end{eqnarray}
%%%%%%%%%
The ingredients of {this Hamiltonian and the} triangular optical lattice~\cite{becker-10} can be engineered in the laboratory with the currently available techniques{~\cite{mandel-03,lee-07,jotzu-15,goldman-10,anderson-13} as stated above}. The frustrated magnetism of the Hamiltonian~(\ref{hamiltonian}) for dominant interactions ($U/t_{\sigma}\gg 1$)~\cite{kuklov-03} is described to leading order in $t_{\uparrow,\downarrow}/U$ by
\begin{eqnarray}
\hat{\mathcal{H}}=\sum_{\langle i,j\rangle}\Big(J\hat{S}_i^x\hat{S}_j^x+J\hat{S}_i^y\hat{S}_j^y+J_z\hat{S}_i^z\hat{S}_j^z\Big)-H\sum_{i}\hat{S}^x_{i},\label{hamiltonianmag}
\end{eqnarray}
where $\hat{S}_i^\alpha=\sum_{\sigma\sigma^\prime} \hat{c}_{i\sigma}^\dagger\bm{\sigma}^\alpha_{\sigma\sigma^\prime}\hat{c}_{i\sigma^\prime}/2$ with the Pauli matrices $\bm{\sigma}^\alpha$ plays the role of quantum spin $S=1/2$. The anisotropy in the spin exchange interactions $J=4t_\uparrow t_\downarrow/U$ and $J_z=2(t_\uparrow^2 + t_\downarrow^2)/U$ must be of easy-axis type ($J\leq J_z$). 
The control of the Rabi frequency translates into tuning a transverse magnetic field, $H=-\Omega$~{\cite{jaksch-05}. Note that the synthetic spin Hamiltonian~(\ref{hamiltonianmag}) has no $U$(1) spin-rotational symmetry due to $\sum_i[\hat{S^\alpha_i},\hat{\mathcal{H}}]\neq 0$ ($\alpha=x,y,z$). }
\begin{figure*}[t]
\includegraphics[scale=0.87]{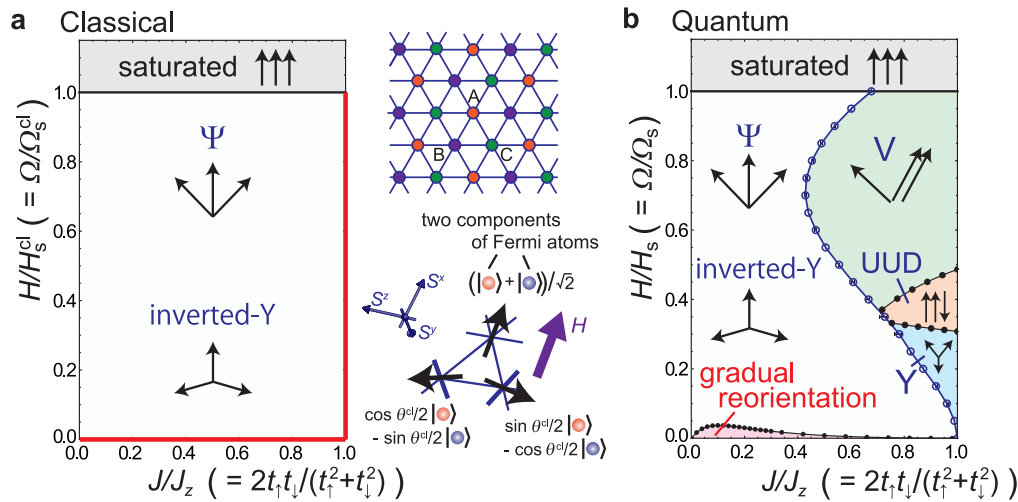}
\caption{\label{fig1}
{\bf Ground-state phase diagrams.} Quantum-fluctuation effects on the ground-state magnetic phases [(b)] in comparison with the classical phase diagram [(a)]. The curves with open (filled) circles indicate first- (second-) order transitions. In (a), there are classical degeneracies of the ground state along {the lines of $J/J_z=1$ and of $H=0$ (red lines; see text)}. The three-sublattice $\sqrt{3}\times\sqrt{3}$ structure and the classical inverted-Y (=$\Psi$) state are illustrated with the mapping between the {representations of spins and two-component Fermi atoms, namely $(\theta,\phi) \leftrightarrow \cos (\theta/2) |\uparrow\rangle +e^{i\varphi} \sin (\theta/2)  |\downarrow\rangle$.}
}
\end{figure*}
%%
% 
%%%%%%%%%%%
%%%%%%%%%%%%
\\\\
{\bf Classical spins.}  Let us begin with a classical-spin analysis as reference to be compared with the quantum case. By treating the spins as classical vectors $\bm{S}_i=(\sin \theta_i\cos \varphi_i,\sin \theta_i\sin \varphi_i,\cos \theta_i)/2$, one gets the classical (mean-field) energy of the system~(\ref{hamiltonianmag}). The minimization of the classical energy with respect to the spin angles $\theta_i$ and $\varphi_i$ leads to a three-sublattice $\sqrt{3}\times\sqrt{3}$ coplanar order in which the polar angles on the sublattices $\mu=A$, $B$, and $C$ are given by $\theta_A=\pi/2$, $\theta_B=\pi-\theta_C=\theta^{\rm cl}\equiv \arcsin \frac{|2H/3-J|}{J+J_z}$ while the azimuthal angles are $\varphi_A=0$ and $\varphi_B=\varphi_C=\pi$ for $0<H<3J/2$ and $\varphi_A=\varphi_B=\varphi_C=0$ for $3J/2<H<H_{\rm s}^{\rm cl}\equiv 3J+3J_z/2$. The ground state has a six-fold degeneracy corresponding to the exchange of the sublattice indices $A$, $B$,  $C$ and resulting {in} a $S_3$ symmetry breaking. As illustrated in Fig.~\ref{fig1}a, the sublattice magnetic moments form an inverted-Y (resp. $\Psi$) shape for low (resp. high) magnetic fields. Note, however, that the apparent change in the expressions for $\theta_\mu$ and $\varphi_\mu$ arises only {from} following the convention $0\leq\theta\leq\pi$ and $0\leq\varphi<2\pi$. 
The inverted-Y and $\Psi$ states are continuously connected and thus equivalent. Therefore, the classical ground state experiences no phase transition up to the magnetic saturation at $H=H_{\rm s}^{\rm cl}$ ($\Omega=\Omega_{\rm s}^{\rm cl}\equiv -3J-3J_z/2$), providing a rather featureless phase diagram (Fig.~\ref{fig1}a).
%%%%%%%%%%%%%%
%%%%%%%%%%%%%%%
 \\ \\
{\bf Quantum phase diagram.} Next the numerical CMF+S method~\cite{yamamoto-12-2,yamamoto-14,yamamoto-17} with 2D DMRG solver is used to unveil quantum fluctuation effects on the classical ground state. We consider a triangular-shaped cluster of $N_C$ spins placed in the environment of self-consistent mean fields $\left(Jm^x_\mu, Jm^y_\mu,J_zm^z_\mu\right)$ (see Methods). The sublattice magnetic moments $m_\mu^\alpha$ characterize the magnetic order. The CMF+S calculations permit the systematic inclusion of quantum-fluctuation effects by increasing $N_C$, which connects between the classical ($N_C=1$) and {exactly quantum} ($N_C\rightarrow \infty$) regimes. 
We will demonstrate that the use of the 2D DMRG solver instead of the Lanczos diagonalization employed in the previous works~\cite{yamamoto-12-2,yamamoto-14,yamamoto-17} {considerably} enlarges the {tractable} cluster size $N_C$, providing the convergence of results for $N_C$ or a systematic series of numerical data for a quantitative extrapolation to $N_C\rightarrow \infty$.

\begin{figure}[t]
\includegraphics[scale=0.45]{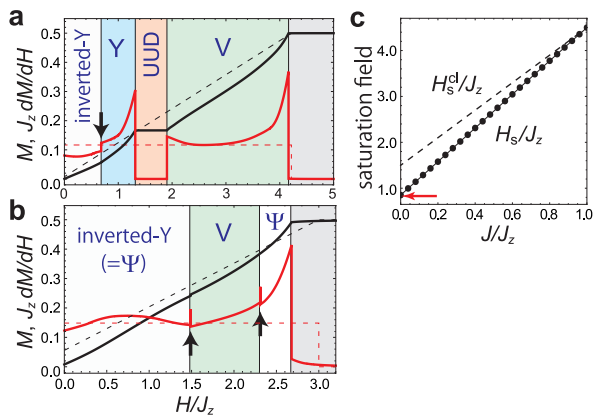}
\caption{\label{fig2}
{\bf Magnetization curves and saturation fields.} The magnetization curve $M=(m^x_A+m^x_B+m^x_C)/3$ and its field derivative $J_z dM/dH$ are plotted for (a) $J/J_z=0.906$ and (b) $J/J_z=0.5$. The quantum CMF+S results (solid lines) are compared to the classical-spin analysis (dashed lines; no phase transition down to $H=H_{\rm s}^{\rm cl}$). The arrows indicate the locations of first-order transitions. (c) Quantum ($H_{\rm s}$) and classical ($H_{\rm s}^{\rm cl}$) values of the saturation field strength as a function of $J/J_z$. The red arrow in (c) marks the previous estimation given in Ref.~\onlinecite{isakov-03} for the transverse Ising model ($J/J_z=0$).} 
\end{figure}
The quantum phase diagram obtained by the CMF+S analysis is shown in Fig.~\ref{fig1}b, which surprisingly reveals various ground states that are absent in the classical counterpart shown in Fig.~\ref{fig1}a. First, we see that the sequence of the three magnetic phases -- coplanar Y-shaped state, collinear up-up-down (UUD) state, and coplanar V-shaped state -- is stabilized from low to high fields for $J/J_z\gtrsim 0.43$. This quantum stabilization has been studied for the isotropic ($J/J_z=1$) case~\cite{chubukov-91,honecker-04,sakai-11}, in which there is an accidental continuous degeneracy of the classical ground-state manifold along the line of $J/J_z=1$ up to the saturation field~\cite{kawamura-85}. The correction to the energy due to the zero-point quantum fluctuations selects the Y-UUD-V sequence from the large ground-state manifold~\cite{chubukov-91}, resulting in the formation of a plateau structure in the magnetization curve, corresponding to the UUD phase with one third of the saturation magnetization~\cite{chubukov-91,honecker-04,sakai-11}. Our quantum phase diagram in Fig.~\ref{fig1}b shows that this quantum effect actually spreads over a wide region of $J/J_z$, especially for high magnetic fields. Note that the UUD magnetization plateau is not exactly flat for $J/J_z<1$ since the $U(1)$ spin rotational symmetry is absent, even though the slope is 
%still 
{quite} small (see Fig.~\ref{fig2}a). For the same reason, the saturation field $H_{\rm s}$ is affected by quantum fluctuations and reduced from the classical value $H_{\rm s}^{\rm cl}$ except at $J/J_z=1$, as shown in Fig.~\ref{fig2}c.
The maximum extent of the V phase in $J/J_z$ is reached at $H/H_{\rm s}\approx 0.71$ ($0.43\lesssim J/J_z\leq 1$), and thus the magnetization process exhibits a reentrant behavior for $0.43\lesssim J/J_z\lesssim 0.68$ from the inverted-Y  to V and back to $\Psi$ state (which is continuously connected to the inverted-Y state) as shown in Fig.~\ref{fig2}b.

The CMF+S phase boundaries have been determined by extrapolating the results of $N_C=15$, $21$, and $36$ (see Methods {for details}). The error bars (shown in Fig.~\ref{fig1}b) estimated from the variation in the linear fittings for different pairs of the $N_C=15,21,36$ data are smaller than the symbol size only with a few exceptions. As an example {of validation}, the CMF+S value for the saturation field at $J/J_z=0$ ($H_s/J_z\approx 0.85$) is in {good} agreement with the previous value, $H_s/J_z= 0.825\pm 0.025$, obtained by the QMC simulations~\cite{isakov-03}.
%%%%%%%%%%%%%%%%%
%%%%%%%%%%%%%%%%%
%%%%%%%%%%%%%
%%%%%%%%%%%%%
%
%%
\begin{figure}[t]
\includegraphics[scale=0.41]{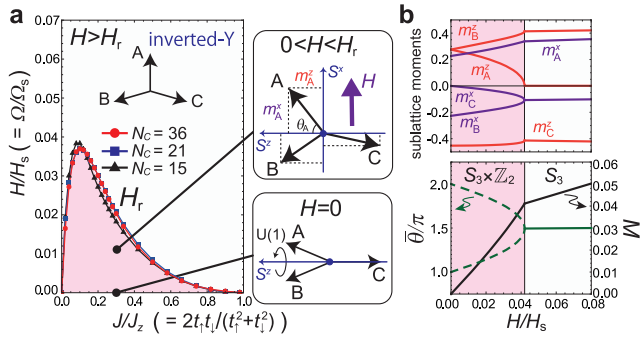}
\caption{\label{fig3}
{\bf Novel spin reorientation transition.} (a) Enlarged view of the low-field region of Fig.~\ref{fig1}b. The quantities characterizing the reorientation along $J/J_z=0.3$ are plotted in {(b)}. The sublattice indices $A$, $B$, and $C$ can be exchanged.} 
\end{figure}
\\ \\{\bf Fluctuation-driven spin reorientation.} 
Even more interestingly, we find the emergence of a novel spin-reorientation transition in the low-field region of the quantum phase diagram (see Fig.~\ref{fig3}). This stems from a quantum order-by-disorder selection from another nontrivial degenerate manifold of the classical ground state at $H=0$. While the minimization of the classical energy leads to a coplanar ground state, {whose plane is identified by an arbitrary azimuthal angle $\varphi$,} 
only two independent constraints are imposed for the polar angles $\theta_A$, $\theta_B$, and $\theta_C$. 
Hence the classical ground state forms a large degenerate manifold~\cite{miyashita-85,miyashita-86,sheng-92}, {namely} $ S^1 \times S^1$. Quantum fluctuations select 
a specific Y-shape order in which one sublattice spin moment is directed parallel to the easy axis, as illustrated in Fig.~\ref{fig3}a; this is in agreement with the prediction of the linear spin-wave approximation~\cite{sheng-92}.

\begin{figure*}[t]
\includegraphics[scale=0.7]{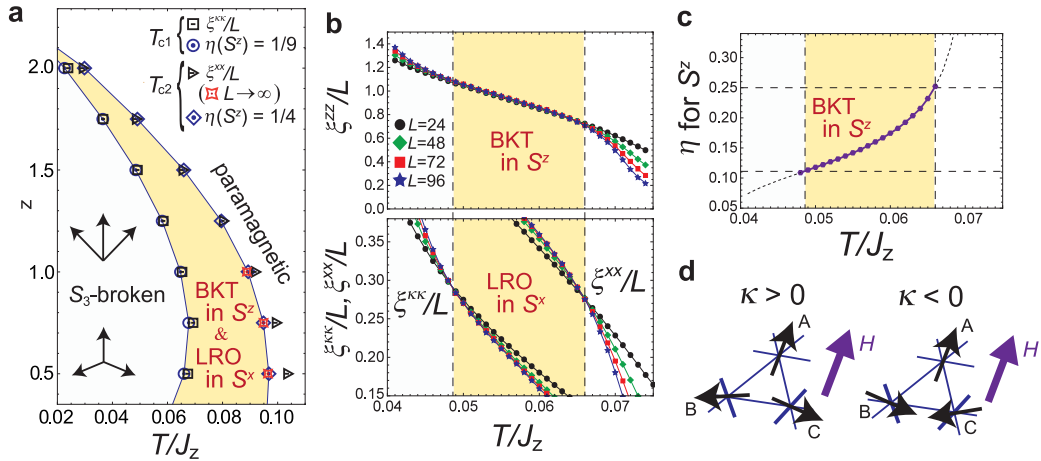}
\caption{\label{fig4}
{\bf Thermodynamic phase transitions.} (a) Thermal phase diagram for $J/J_z=0.3$, obtained by the {Monte Carlo} simulations. The phase boundaries are determined by the quantities in (b) and (c) (the values at $H/J_z=1.5$ are shown). The crossing points in $\xi^{xx}/L$ with $L=24$ to $96$ are slightly different from $T_{{\rm c}2}$ determined from $\eta(S^z)=1/4$ for $H/J_z\lesssim 1$. This is because the former has a strong finite-size effect, and is improved by taking $L\rightarrow\infty$. (d) The inverted-Y states with positive and negative values of chirality.}
\end{figure*}
When a transverse field is applied, a generic three-sublattice state breaks a $S_3\times \mathbb{Z}_2$ {symmetry} (corresponding to the permutation of sublattice indices and $\pi$ rotation around the field axis in  spin space). 
Our CMF+S analysis shows that a quantum magnetic phase that indeed breaks {such symmetry completely} is realized with a coplanar spin order in the $S^z$-$S^x$ plane in a low-field region of the quantum phase diagram. However, when the magnetic field strength $H$ exceeds a threshold value $H_{\rm r}$, the classical inverted-Y phase (which breaks $S_3$ symmetry, as mentioned above)  is recovered, with a quantum reduction in the length of the ordered moments and a slight change in the relative angles. Consequently, the structure of the sublattice moments is gradually changed from a Y shape to another Y shape with different orientations via the intermediate $S_3\times \mathbb{Z}_2$-broken phase (see Fig.~\ref{fig3}b), which does not occur in the classical case. The behavior of this reorientation transition can be characterized by a parameter $\bar{\theta}\equiv \theta_A+\theta_B+\theta_C$, which is independent of the sublattice-exchange operation. As shown in Fig.~\ref{fig3}c, the value of $\bar{\theta}$  gradually changes from $\pi$ (or $2\pi$) to $3\pi/2$ when $H$ is increased from 0 to $H_{\rm r}$. The two branches, one of which is spontaneously chosen, correspond to the remaining $\mathbb{Z}_2$ degeneracy with respect to the $\pi$ rotation around the field axis. 
%%%%%%%%%%%%%%
%%%%%%%%%%%%%%
\\ \\ {\bf Thermodynamic phase diagram.} Finally, let us explore novel thermal-fluctuation effects in this frustrated system {and}  provide the estimation of the ordering transition temperature to assist {with} the experimental implementation. Since the QMC simulations suffer from the minus-sign {problem~\cite{suzuki-93}, we replace the spin operators in the Hamiltonian~(\ref{hamiltonianmag}) with classical spin vectors $\bm{S}_i$ of length $1/2$ and} perform {Monte Carlo} simulations on $L=24,48,72,96$ under periodic boundary conditions (see Methods). The classical {Monte Carlo} analysis is expected to provide a reasonable {approximation} in the temperature range away from the low-temperature quantum region. Since the stabilization of the Y-UUD-V sequence due to the thermal fluctuations has been studied for the isotropic ($J/J_z=1$) model~\cite{kawamura-85,seabra-11}, here we focus on the paramagnetic transition from the inverted-Y ($=\Psi$) phase for a relatively small $J/J_z$.

The thermal phase diagram for $J/J_z=0.3$ is summarized in Fig.~\ref{fig4}a. Supposing the inverted-Y order, we estimate the correlation length for the three-sublattice order in the $S^z$ and $S^x$ spin components ($\xi^{zz}$, $\xi^{xx}$) and for the uniform order in the $y$ component of the vector chirality $\kappa$ ($\xi^{\kappa\kappa}$). The scaled quantity $\xi^{\alpha\alpha}/L$ should become independent of $L$ at a critical point. 
As shown in Fig.~\ref{fig4}b, $\xi^{zz}/L$ exhibits a finite critical range in $T/J_z$ with a constant value independent of $L$, which indicates that a BKT phase with quasi-LRO appears despite the absence of continuous symmetry in the model~(\ref{hamiltonianmag}). In the critical range, $T_{{\rm c}1}<T<T_{{\rm c}2}$, the $S^z$ correlation function exhibits algebraic decay $\sim |\bm{r}_i-\bm{r}_j|^{-\eta}$ with {critical} exponent $\eta$. 
This can be explained from the discrete but high degree  (six-fold)  of degeneracy of the inverted-Y ground state, given the fact that {two-dimensional systems} with the discrete $\mathbb{Z}_p$ symmetry are believed to exhibit a BKT behavior for $p>4$~\cite{jose-77}. Figure~\ref{fig4}c shows the exponent $\eta$ for $S^z$ correlations, which takes values $\sim 1/9$ and $\sim 1/4$ at the lower ($T_{{\rm c}1}$) and upper ($T_{{\rm c}2}$) endpoints of the critical region. This is indeed consistent with the expectation for the universality of the 2D 6-state clock model~\cite{jose-77}, whose $\mathbb{Z}_6$ symmetry group has six elements, same as the $S_3$ symmetry of the present case. 
%%%%%%%%%%%%%%
%%%%%%%%%%%%%%
\\ \\ {\bf Coexistence of BKT and LRO.} 
Interestingly, however, $\xi^{xx}/L$ and $\xi^{\kappa\kappa}/L$ show an isolated critical point, unlike $\xi^{zz}/L$, as seen in Fig.~\ref{fig4}b as the crossings of the curves for different $L$. The critical points for $S^x$ and $\kappa$ are coincident or very close to the upper and lower endpoints of the BKT behavior in $S^z$, respectively. Consequently, in the intermediate phase for $T_{{\rm c}1}<T<T_{{\rm c}2}$, BKT-type quasi-LRO in the field-perpendicular component ($S^z$) and LRO in the field-parallel spin component ($S^x$) coexist while the chirality is disordered. This particular coexistence of BKT and LRO should be attributed to the difference of $S_3$ from $\mathbb{Z}_6$; the nonabelian {symmetric} group $S_3$ is formed by the cycles and transpositions of $\{A,B,C\}$ and has two generators, while $\mathbb{Z}_6$ is just formed by the cycles of six elements and has one generator. In fact, the breaking of {the} even and odd permutation symmetries is detected by $S^x$ and $\kappa$, respectively. The $\mathbb{Z}_6$ clock model instead cannot have such partial symmetry breaking transitions. {Note} that the $S^x$ component of the magnetic moments in the inverted-Y order takes the same value on two of the three sublattices, and thus all exchanges generating three elements can be obtained only by cyclic permutations of the sublattice indices while the chirality changes its sign by transposing two sublattice indices as illustrated in Fig.~\ref{fig4}d.
%%%%%%%%%%
%%%%%%%%%%
%
%%
\begin{figure*}[t]
\includegraphics[scale=0.75]{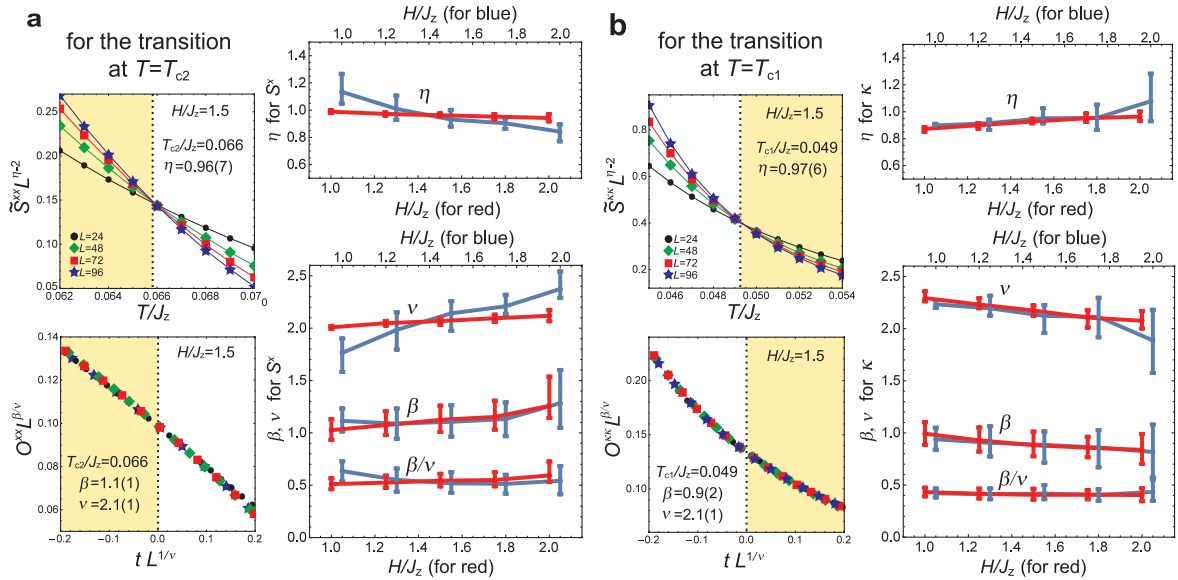}
\caption{\label{fig5}
{\bf Scaling analyses and critical exponents.} The scaling analyses on the structure factor $\tilde{\mathcal{S}}^{\alpha\alpha}$ at the appropriate wave vector and the order parameter $\mathcal{O}^{\alpha\alpha}$ (see Methods) for the transitions (a) at $T_{{\rm c}2}$ ($\alpha=x$) and (b) at $T_{{\rm c}1}$ ($\alpha=\kappa$) are shown for $H/J_z=1.5$. The scaled temperature is defined by $t\equiv (T-T_{{\rm c}1,2})/T_{{\rm c}1,2}$. The extracted critical exponents are plotted as a function of $H/J_z$. The blue and red symbols are the values obtained by using the critical points $T_{{\rm c}2}$ ($T_{{\rm c}1}$) determined by the crossings in the scaled correlation length $\xi^{xx}/L$ ($\xi^{\kappa\kappa}/L$) and by the condition $\eta=1/4$ ($\eta=1/9$) for the $S^z$ correlation function, respectively. The scales of the upper and lower horizontal axes for the blue and red data points are slightly shifted {from} each other for the {sake of} visibility.}
\end{figure*}
\\ \\ {\bf Possible new universality class.} 
It is noteworthy that the critical exponents associated with diverging $\xi^{xx}$ and $\xi^{\kappa\kappa}$ have no correspondence in the known universality classes. From the scaling analyses (see Methods), we extract the values of the correlation-function ($\eta$), order-parameter ($\beta$), and correlation-length ($\nu$) exponents for the transitions in $S^x$ at $T_{{\rm c}2}$ and in $\kappa$ at $T_{{\rm c}1}$, respectively (see Fig.~\ref{fig5}). In general, the estimation of critical exponents are very sensitive to the determination of critical points. We showed the critical exponents extracted with two different {ways to estimate}  $T_{{\rm c}2}$ ($T_{{\rm c}1}$), {namely} from the crossings in the scaled correlation length $\xi^{xx}/L$ ($\xi^{\kappa\kappa}/L$) and from the condition $\eta=1/4$ ($\eta=1/9$) for the $S^z$ correlation function. Note that the latter estimation of the critical points has less finite-size effect. 
As shown in Fig.~\ref{fig5}a, the values of $\eta$ and $\beta/\nu$ for $S^x$ are roughly constant (at $\sim 1$ and $\sim 0.5$) for varying $H/J_z$, keeping the {2D} hyperscaling law $\eta=2\beta/\nu$, which {indicates} the existence of a new {weak} universality~\cite{suzuki-74}. 
Moreover, the individual exponents $\beta$ and $\nu$ might also be constant (implying {a new universality in the usual sense}), although we {cannot reach} a final conclusion with the given numerical data. Note that {by trying several values of  $J/J_z$, we could also confirm that the exponents are roughly independent of the spin anisotropy,  to the same degree as they are of the magnetic field $H/J_z$.  The critical exponents associated with the chiral transition at $T_{{\rm c}1}$ also have} a similar behavior as shown in Fig.~\ref{fig5}b. In {both transitions at $T_{{\rm c}2}$ and $T_{{\rm c}1}$, the critical exponents are clearly distinguished from the naively expected 2D three-state Potts ($\eta=4/15$, $\beta=1/9$, $\nu=5/6$) and Ising ($\eta=1/4$, $\beta=1/8$, $\nu=1$) ones,} {which means that the two-step transition and the emergence of the intermediate phase with the BKT and LRO coexistence are {indeed a new phenomenon and not a simple sequence of standard $\mathbb{Z}_3$ and $\mathbb{Z}_2$ symmetry breaking transitions.} 
%%%%%%%%%%% 
%%%%%%%%%%%
\\ \\    
{\bf \large Discussion}\\
We studied the quantum and thermal phase transitions in the easy-axis triangular XXZ model under transverse magnetic fields and proposed {their} quantum simulation with coherently-coupled binary gases of ultracold fermionic atoms. In the cold-atom Hubbard simulator, the spin-dependent hoppings and coherent coupling between the two components play the role of the spin exchange anisotropy and transverse magnetic fields, respectively. We found that in the ground state, the order-by-disorder effects {induced} by the quantum fluctuations {give} rise to several nontrivial phase transitions, including a novel spin reorientation. Although the region of the reorientation transition is limited to low magnetic fields ($H\lesssim 0.04H_{\rm s}$), this poses a minor threat for future solid-state experiments, {in which} the saturation field $H_{\rm s}$ is usually very large. The reorientation could be observed as a kink in the magnetization curve or, more conclusively, in the inelastic neutron scattering. The detection in cold-atom quantum simulators should be more challenging since a precise tuning of Rabi frequency $\Omega$ and }{a sufficiently} low temperature are required.

Moreover, the  transition from the paramagnetic to the ordered state with $S_3$ symmetry breaking was found to exhibit a particular two-step {behavior} with an intermediate phase that showed a coexistent BKT and LRO {correlations}. The temperature scale of the transitions was estimated from the {Monte Carlo} analysis to be at most of order $0.1J_z$ (see Fig.~\ref{fig4}a), with little sensitivity to the variation of $J/J_z$. With the caveat that quantum fluctuations can actually lower this  temperature to some extent, this estimation should serve as a guideline for future developments of cold-atom quantum simulators for studying novel frustrated magnetism.

Also {from the standpoint of more fundamental statistical physics,} our finding of the possible new (weak) universality class related to the $S_3$ symmetry breaking should stimulate a further investigation of phase transition phenomena {with a special focus on the comparison between the well-studied cyclic groups $\mathbb{Z}_p$ and the other symmetry} groups with the same order, such as $\mathbb{Z}_4$ versus $V_4\simeq \mathbb{Z}_2\times \mathbb{Z}_2$ and $\mathbb{Z}_{10}\simeq \mathbb{Z}_5\times \mathbb{Z}_2$ versus the nonabelian $D_{10}$; this will build a deeper understanding of the spontaneous breaking of discrete symmetries. {It is {important} to remark that the Ising limit ($J=0$) of our model~(\ref{hamiltonianmag}) has been simulated very recently with the D-Wave quantum annealing processor~\cite{king-2018}, which has a great potential to test the partial discrete symmetry breaking transitions that we found and the associated critical phenomena. }

\begin{figure}[t]
\includegraphics[scale=0.84]{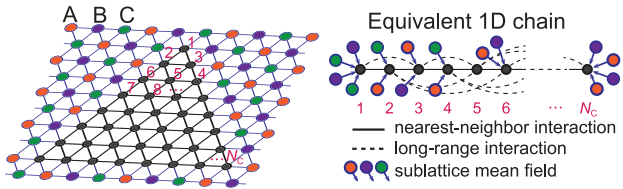}
\caption{\label{fig6}
{\bf 2D DMRG under mean-field boundary conditions.} Finite-size 2D cluster embedded in the background three-sublattice structure, which is mapped onto an equivalent 1D chain with long-range interactions and mean fields.  }
\end{figure}

\begin{figure}[t]
\includegraphics[scale=0.45]{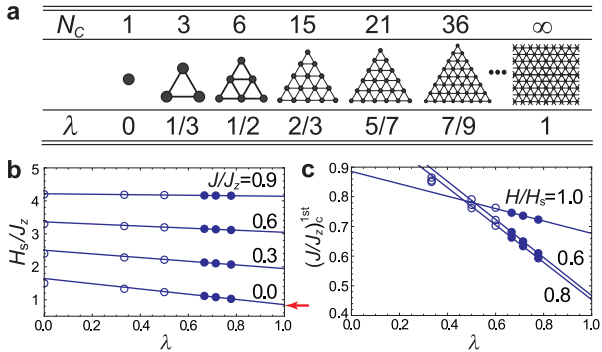}
\caption{\label{fig7}
{\bf The CMF+S analysis.} (a) Series of the clusters of $N_C$ spins with the scaling parameter $\lambda$. Examples of the CMF+S scalings, which determine the quantum phase diagram in Fig.~\ref{fig1}b, for (b) the saturation field and (c) the first-order transition point between the inverted-Y (=$\Psi$) and V phases. The red arrow in (b) marks the previous estimation given in Ref.~\onlinecite{isakov-03} for the transverse Ising model.} 
\end{figure}

{Finally}, we {emphasize} that our procedure for developing the CMF+S scheme with the use of a 2D DMRG solver significantly expands the scope of its applications to {broader} areas of quantum frustrated systems. For example, the extension {of the cluster size $N_C$ enables us to deal with more challenging and physically rich lattice systems such as the ones on the kagome lattice,} which requires a series of larger-size clusters because of the geometry. 
%%%%%%%%%%% 
%%%%%%%%%%%
\\ \\    
{\bf \large Methods}\\
{\bf The CMF+S with 2D DMRG solver.} We develop the numerical CMF+S method~\cite{yamamoto-12-2,yamamoto-14,yamamoto-17} by employing the 2D DMRG  as a solver of the finite-size cluster problem. We consider a triangular-shaped cluster of $N_C$ spins, in which the interaction between a cluster-edge spin and its neighboring spin at an out-of-cluster site with sublattice index $\mu$ is replaced by an effective magnetic field $\left(Jm^x_\mu, Jm^y_\mu,J_zm^z_\mu\right)$ acting on the edge spin (see Fig.~\ref{fig6}). The sublattice magnetic moment $m_\mu^\alpha$, which is self-consistently determined via the condition $m_\mu^\alpha\equiv \frac{3}{N_C}\sum_{i_\mu}^{N_C/3}\langle \hat{S}_{i_\mu}^\alpha \rangle$, characterizes the magnetic phase. We employ a series of clusters with $N_C=3,6,15,21,36$ listed in Fig.~\ref{fig7}a to carry out an efficient extrapolation $N_C\rightarrow \infty$. As a solver for each cluster problem to obtain the expectation value $\langle \hat{S}_{i_\mu}^\alpha \rangle$, we use the DMRG {method} in order to overcome the past practical system-size limitation of the Lanczos solver~\cite{yamamoto-14,yamamoto-17}. The DMRG method, {whose efficiency is well-known for 1D systems,} can be applied to the 2D cluster problem by mapping it to an equivalent 1D chain with long-range interactions as shown in Fig.~\ref{fig6}. For our CMF+S analyis, in addition, the effective magnetic fields coming from the mean-field boundary conditions for the original 2D cluster have to be treated on each site. Therefore, in order to determine the values of mean fields, $\left(Jm^x_\mu, Jm^y_\mu,J_zm^z_\mu\right)$, in a self-consistent way, one has to iteratively perform the DMRG calculations until convergence. The dimension of the truncated matrix product states in the DMRG is kept sufficiently large (typically $\sim 10^3$-$10^4$) to obtain numerically precise values of $\langle \hat{S}_{i_\mu}^\alpha \rangle$. For the tensor operations, we used the ITensor package~\cite{ITensor}. {It is also worth mentioning the CMF analysis of Ref.~\onlinecite{suzuki-14} on a stacked layered system, in which an effective cluster problem on a three-leg tube along the stacking direction has been treated with the DMRG method. }

The CMF+S phase boundaries in Fig.~\ref{fig1}b were determined by extrapolating the results of $N_C=15$, $21$, and $36$ with a {simple} linear function of the scaling parameter $\lambda\equiv N_B/3N_C$ with $N_B$ being the number of bonds within the cluster. We present the examples of the extrapolations in Fig.~\ref{fig7}b and~\ref{fig7}c, which show that the  linear fittings {actually}  work well. 
%%
%
%%%%%%%%%%% 
%%%%%%%%%%%
\\ \\
{\bf Details of {Monte Carlo} analysis.} To discuss the thermal fluctuation effects, we perform the {Monte Carlo} simulations on $L\times L$ rhombic clusters with $L=24,48,72,96$ under periodic boundary conditions. Typical simulations contain $2\times 10^6$ MC steps, each of which consists of one Metropolis sweep followed by two over-relaxation sweeps. To discuss the spontaneous symmetry breaking, we estimate the correlation length for the $\alpha$ component of spins
\begin{eqnarray}
\xi^{\alpha\alpha}=\frac{\sqrt{3}L}{4\pi}\sqrt{\frac{\mathcal{S}^{\alpha\alpha}(\tilde{\bm{k}})}{\mathcal{S}^{\alpha\alpha}(\tilde{\bm{k}}+(0,4\pi/\sqrt{3}L))}-1}
\end{eqnarray}
from the structure factor
\begin{eqnarray}
\mathcal{S}^{\alpha\alpha}(\bm{k})=\frac{1}{L^2}\sum_{i,j}\langle S_i^\alpha S_j^\alpha\rangle e^{-i\bm{k}\cdot(\bm{r}_i-\bm{r}_j)}.
\end{eqnarray}
Similar quantities are also defined for the $y$ component of the vector chirality $\kappa_i\equiv -\frac{2}{3\sqrt{3}}(\bm{S}_i\times \bm{S}_{i+\bm{a}_1}+\bm{S}_{i+\bm{a}_1}\times \bm{S}_{i+\bm{a}_2}+\bm{S}_{i+\bm{a}_2}\times \bm{S}_{i})_y$ with $\bm{a}_{1,2}$ being the primitive vectors of the triangular lattice. Supposing the three-sublattice inverted-Y order, we take $\tilde{\bm{k}}=(4\pi/3,0)$ for the $S^z$ and $S^x$ components and $\tilde{\bm{k}}=(0,0)$ for the chirality.

In order to extract the critical exponents $\eta$, $\beta$, and $\nu$, we define the quantities $\tilde{\mathcal{S}}^{\alpha\alpha}\equiv \mathcal{S}^{\alpha\alpha}(\tilde{\bm{k}}) /(T/J_z)$ and $\mathcal{O}^{\alpha\alpha}\equiv \sqrt{\mathcal{S}^{\alpha\alpha}(\tilde{\bm{k}})}/L$. 
Those quantities obey the following scaling relations:
\begin{eqnarray}
\tilde{\mathcal{S}}^{\alpha\alpha}&=&L^{2-\eta} \tilde{\mathcal{S}}^{\alpha\alpha}_0 (t L^{1/\nu}),\nonumber\\
\mathcal{O}^{\alpha\alpha}&=&L^{-\beta/\nu} \mathcal{O}^{\alpha\alpha}_0 (t L^{1/\nu})
\end{eqnarray}
with unknown universal functions $\tilde{\mathcal{S}}^{\alpha\alpha}_0$ and $\mathcal{O}^{\alpha\alpha}_0$ of $t=(T-T_{{\rm c}1,2})/T_{{\rm c}1,2}$. We determine the exponents $\eta$ by the condition that the scaled quantity $\tilde{\mathcal{S}}^{\alpha\alpha}/L^{2-\eta}$ should become independent of system size at $T=T_{{\rm c}1,2}$. The exponents $\beta$ and $\nu$ are determined through the data collapse for different $L$ in $\mathcal{O}^{\alpha\alpha} L^{\beta/\nu}$ as a function of $t L^{1/\nu}$ around the critical point (see Fig.~\ref{fig5}).

%%%%%%%%%%% 
%%%%%%%%%%%
{\bf \large Acknowledgements}\\
{This work was supported by KAKENHI from Japan Society for the Promotion of Science, Grant Numbers 18K03525 (D.Y.), 18K03492 (I.D.), 18H05228 (I.D.), and CREST from Japan Science and Technology Agency No.~JPMJCR1673. }
G.M. wants to thank P.~Calabrese and E.~Vicari for useful correspondence.
G.M. is supported by the Ministry of Education, Culture, Sports, Science (MEXT)-Supported Program for the Strategic Research Foundation at Private Universities ``Topological Science'' (Grant No. S1511006). 
Part of the calculations was carried out using the TSC computer of the ``Topological Science'' project in Keio University.

\end{document}